\documentclass[twocolumn,10pt,cleanfoot]{asme2e}

\usepackage{graphicx}
\usepackage{cite}

\confshortname{SMASIS 2018}
\conffullname{the ASME 2018\\ Conference on Smart Materials, Adaptive Structures and Intelligent Systems}

\confdate{September 10-12}
\confyear{2018}
\confcity{San Antonio, TX}
\confcountry{USA}

\papernum{SMASIS2018-8190}

\title{Crack Growth Behavior in NiTi Shape Memory Alloys under Mode-I Isothermal Loading: Effect of Stress State
}

\author{Behrouz Haghgouyan\\
       {\tensfb Ibrahim Karaman}
    \affiliation{
	Department of Materials Science and Engineering\\
	Texas A\&M University\\
	College Station, Texas 77843\\
    }	
}

\author{Sameer Jape\\
       {\tensfb Alexandros Solomou}\\
       {\tensfb Dimitris C. Lagoudas}
    \affiliation{
	Department of Aerospace Engineering\\
	Texas A\&M University\\
	College Station, Texas 77843\\
    }	
}

\begin{document}

\maketitle

\begin{abstract}
{\it Fracture behavior in nickel-titanium (NiTi) shape memory alloys (SMAs) subjected to mode-I, isothermal loading is studied using finite element analysis (FEA). Compact tension (CT) SMA specimen is modeled in Abaqus finite element suite and crack growth under displacement boundary condition is investigated for plane strain and plane stress conditions. Parameters for the SMA material constitutive law implemented in the finite element setup are acquired from characterization tests conducted on near-equiatomic NiTi SMA. Virtual crack closure technique (VCCT) is implemented where crack is assumed to extend when the energy release rate at the crack-tip becomes equal to the experimentally obtained material-specific critical value. Load-displacement curves and mechanical fields near the crack-tip in plane strain and plane stress cases are examined. Moreover, a discussion with respect to the crack resistance R-curves calculated using the load-displacement response for plane strain and plane stress conditions is presented.}
\end{abstract}

\section*{INTRODUCTION}
\label{intro}
Unique properties of Shape Memory Alloys (SMAs), such as pseudoelasticity and shape memory effect, are driven by a reversible, solid-to-solid, diffussionless transformation between austenite and martensite phases, triggered by temperature variations and/or mechanical loading. SMAs have found applications in aerospace, automotive, biomedical, oil and gas exploration, and energy storage industries owing to their superior performance when compared to conventional materials~\cite{lagoudas2008shape}. Studying the failure mechanisms of SMAs has been of recent interest~\cite{mohajeri2018evolution, mohajeri2018nickel, mohajeri2018corrosion, phillips2018effect}. Understanding the fracture response and crack growth behavior of SMAs is thus crucial for robust and reliable design of geometrically large SMA components and their successful integration into smart structures. However, presence of stress-induced transformation zone near the crack-tip, along with nonlinear thermomechanical material response in SMAs (\emph{viz.} transformation induced plasticity, martensite variant reorientation, and strong thermomechanical coupling) complicates the study of their fracture behavior~\cite{baxevanis2015fracture}.

Implementing various constitutive models~\cite{boyd1996thermodynamical, zaki2007three, hartl2012phenomenological, Xu2017trip, xu2017finite}, fracture of SMAs has been studied numerically in the presence of stationary cracks~\cite{yan2003theoretical, wang2005formation,  maletta2010analytical, baxevanis2012finite, baxevanis2012mode, shafaghi2015experimental, hazar2016evaluation, haghgouyan2016exp}. It is shown that stress concentration at the crack tip causes forward transformation (\emph{i.e.} austenite to martensite) that acts as a mechanism for energy dissipation similar to plastic deformation. Energy dissipation at the crack tip through forward phase transformation can lead to stable crack growth under nominally isothermal mechanical loading~\cite{yan2002effect, freed2007crack, creuziger2008fracture, baxevanis2013fracture, hazar2015modeling, kelton2017study, kelton2018visualization, haghshenas2018damage, mehdizadeh2018role, haghgouyan2018SPIE} and thermo-mechanical loading ~\cite{Baxevanis2016, jape2016stable, Iliopoulos2017, jape2017fracture,JBL-14,JBPL-15,JSBL-16}. In the present work, crack growth in SMAs under mode-I, isothermal loading is investigated via finite element analysis (FEA). Compact tension (CT) specimen is modeled in Abaqus where the parameters for the model calibration are acquired from near-equiatomic nickel-titanium (NiTi) SMA. Analysis is carried out for mode-I, isothermal monotonic loading and crack growth response under both plane strain and plain stress conditions is examined.

The remainder of the paper is organized as follows. First, a constitutive material model that describes the thermomechanical response of polycrystalline SMAs is given. Next, formulation of the finite element boundary value problem, solution methodology and fracture criterion are described. Then, crack growth simulation results are presented and the key findings are summarized and their implications discussed.

\section*{MATERIAL MODEL}
\label{model}

In this section, the constitutive model for polycrystalline SMAs proposed by Boyd and Lagoudas \cite{boyd1996thermodynamical} is described. This model is developed within the framework of continuum thermodynamics and adopts the classical rate-independent small-strain flow theory for the evolution equations of the transformation strains.

The increments of the strain tensor components are given by
\begin{equation}\label{dStress}
d\varepsilon_{ij} = S_{ijkl} d\sigma_{kl} + d S_{ijkl}  \sigma_{kl} + d{\varepsilon}^t_{ij},
\end{equation}
where ${S}_{ijkl}$ are the components of the current compliance tensor, and $\sigma_{ij}$, $\varepsilon_{ij}^t$ are the Cartesian components of the stress tensor and transformation strain tensor, respectively. ${S}_{ijkl}$ vary with the martensite volume fraction, $\xi$, as
\begin{equation}
S_{ijkl}=(1-\xi) S^A_{ijkl} + \xi S^M_{ijkl},
\end{equation}
where $S^A_{ijkl}$, $S^M_{ijkl}$ are the components of the austenite and martensite compliance tensor, respectively. Assuming elastic isotropy for both phases
\begin{equation}
S^{\alpha}_{ijkl}=\frac{1+\nu_{\alpha}}{2E_{\alpha}}(\delta_{il}\delta_{jk}+\delta_{ik}\delta_{jl})-\frac{\nu_{\alpha}}{E_{\alpha}}\delta_{ij}\delta_{kl}, \end{equation}
where $\alpha$ stands for $A$ or $M$ for austenite or martensite, respectively. $E$ is Young's modulus, $\nu$ is Poisson's ratio, and $\delta_{ij}$ is Kronecker's delta. The transformation strain, $\varepsilon^t$, is related to the martensitic volume fraction, $\xi$, by the evolution equation
\begin{equation}\label{evolution}
d{\varepsilon}^{t}_{ij}= \Lambda_{ij} d\xi,
\end{equation}
where $\Lambda_{ij}$, the components of the direction tensor, are defined during forward transformation as
\begin{equation}\label{tensor}
\Lambda^{fwd}_{ij}=\frac{3}{2}\frac{H^{cur}}{\bar{\sigma}} s_{ij}.
\end{equation}
Here, $\bar{\sigma}$ is the Mises equivalent stress $\bar{\sigma}=\sqrt{\frac{3}{2}s_{ij}s_{ij}}$ and $s_{ij}$ are the component of deviatoric stress tensor
$s_{ij}=\sigma_{ij}-\sigma_{kk}\delta_{ij}/3$. $H^{cur}$ is the transformation strain at the current stress level and is represented as an exponential function of the applied stress:
\begin{equation}\label{H}
H^{cur}(\bar{\sigma})=H_{sat}\left(1-e^{-k\bar{\sigma}}\right),
\end{equation}
where $H_{sat}$ is saturated value of $H^{cur}$ and $k$ controls the rate at which $H^{cur}$ exponentially evolves from 0 to $H_{sat}$. During forward transformation, the stress remains on the transformation surface, \emph{i.e.} $\Phi^{fwd}=0$, where
\begin{equation}
\Phi^{fwd}=\pi^{fwd}- Y_0, \qquad d\xi>0.
\end{equation}
$\pi^{fwd}$ is the thermodynamic driving forces for forward transformation and $Y_0$ is the threshold value for activation of the transformation. $\pi^{fwd}$ is given by
\begin{equation}
\pi^{fwd}=\sigma_{ij}\Lambda^{fwd}_{ij}+\frac{1}{2}{\Delta S_{ijkl}}\sigma_{ij}\sigma_{kl}+\rho \Delta s_0 T -\rho \Delta u_0 - f^{fwd},
\end{equation}
where $T$ is the temperature, $s_0$ is the specific entropy, $u_0$ is the internal energy, $\rho$ is the density and $\Delta$ denotes the difference in property between the martensitic and the austenitic states. $f^{fwd}$ is the hardening function during forward phase transformation
\begin{equation}
f^{fwd}=\frac{1}{2}\alpha_1(1+\xi^{n_1}-(1-\xi)^{n_2})+\alpha_2,
\end{equation}
where $\alpha_1$, $\alpha_2$, $n_1$, $n_2$ are coefficients that assume real number values. Given these constitutive relations, the following model parameters need to be calibrated: (i) the elastic properties of martensite and austenite, (ii) parameters contained in Eq.~(\ref{H}), and (iii) the model parameters that are characteristic of the martensitic transformation, \emph{i.e.} $\rho \Delta s_0$, $\rho \Delta u_0$, $\alpha_1$, $\alpha_2$, $Y_0$.

\begin{figure}[t]
\begin{center}
\includegraphics[width=1\columnwidth]{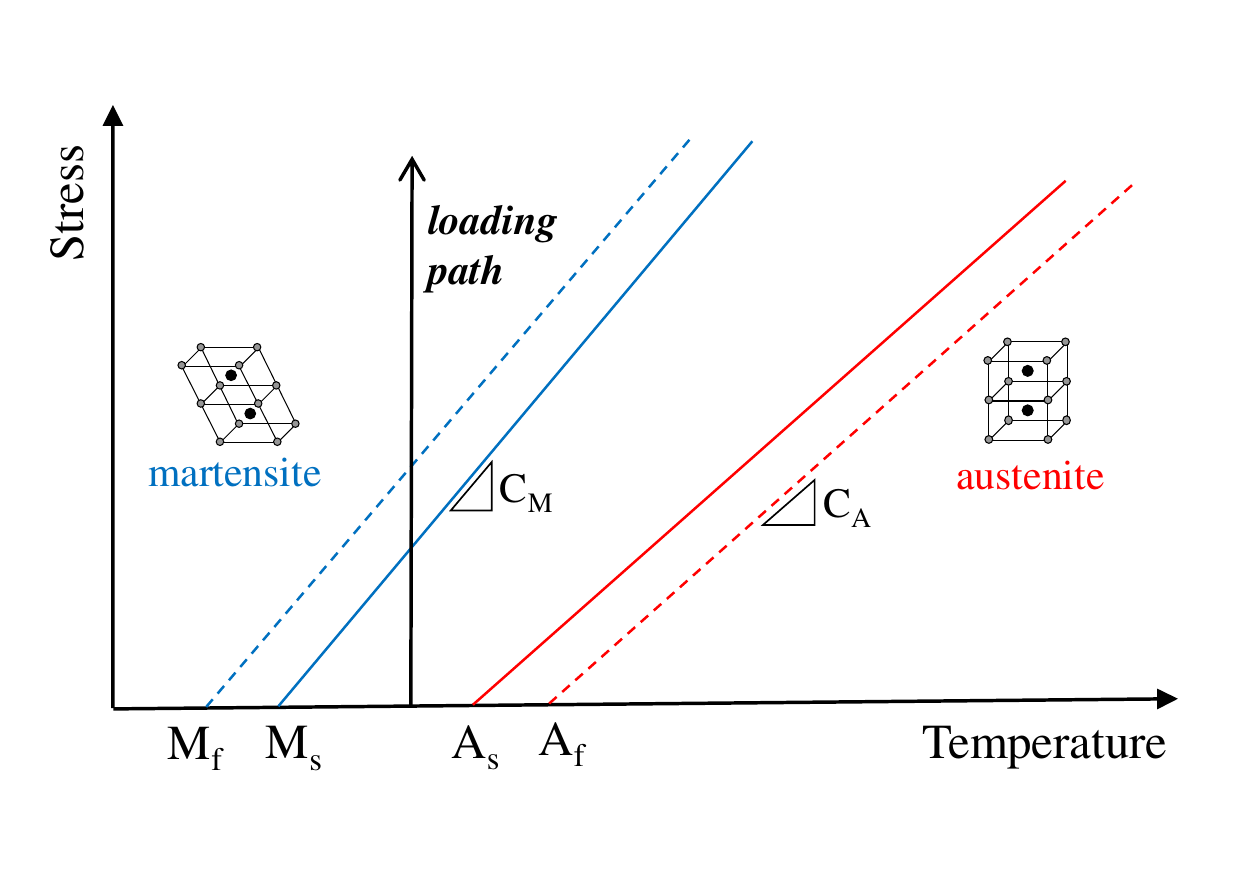}
\end{center}
\caption {Schematic of the phase diagram and loading path.}
\vskip\baselineskip
\label{PATH}
\end{figure}

The material properties that are used to calibrate the model are $E_A$, $E_M$, $\nu_A$, $\nu_M$, $H_{sat}$, $M_s$, $M_f$, and $C_M$. Here, $M_s$, $M_f$ are the martensitic start and martensitic finish temperatures at zero load, respectively, and $C_M$ is the forward transformation slope in the stress--temperature phase diagram (Fig.~\ref{PATH}). The parameters used in this study are shown in Table~\ref{Table1}. The elastic constants are measured from isothermal stress--strain curves. The parameters for $H^{cur}(\bar{\sigma})$ are calibrated using isobaric testing, and the remaining parameters are obtained from the corresponding stress--temperature curves \cite{lagoudas2008shape}.

\begin{table}
\vspace{0mm}
\centering
\begin{tabular}{ l l@{\qquad \qquad} l l}
    \hline
    Parameter & Value   & Parameter & Value \\ \hline
    $E_A$     & 68~GPa  & $M_{s}$   & 340~K \\
    $E_M$     & 62~GPa  & $M_{f}$   & 330~K \\
    $\nu_A$, $\nu_M$ & 0.33    & $C_{M}$   & 15~MPa/K \\
    $H_{sat}$ & 0.05    & $k_{t}$   & 0.026~MPa$^{-1}$ \\ \hline
\label{Table1}
\end{tabular}
\caption[example]
{ \label{Table1}
Material parameters used for finite element simulations.}
\end{table}

\section*{METHODOLOGY}
\label{sec:problem}
\begin{figure}[t]
\begin{center}
\includegraphics[width=1\columnwidth]{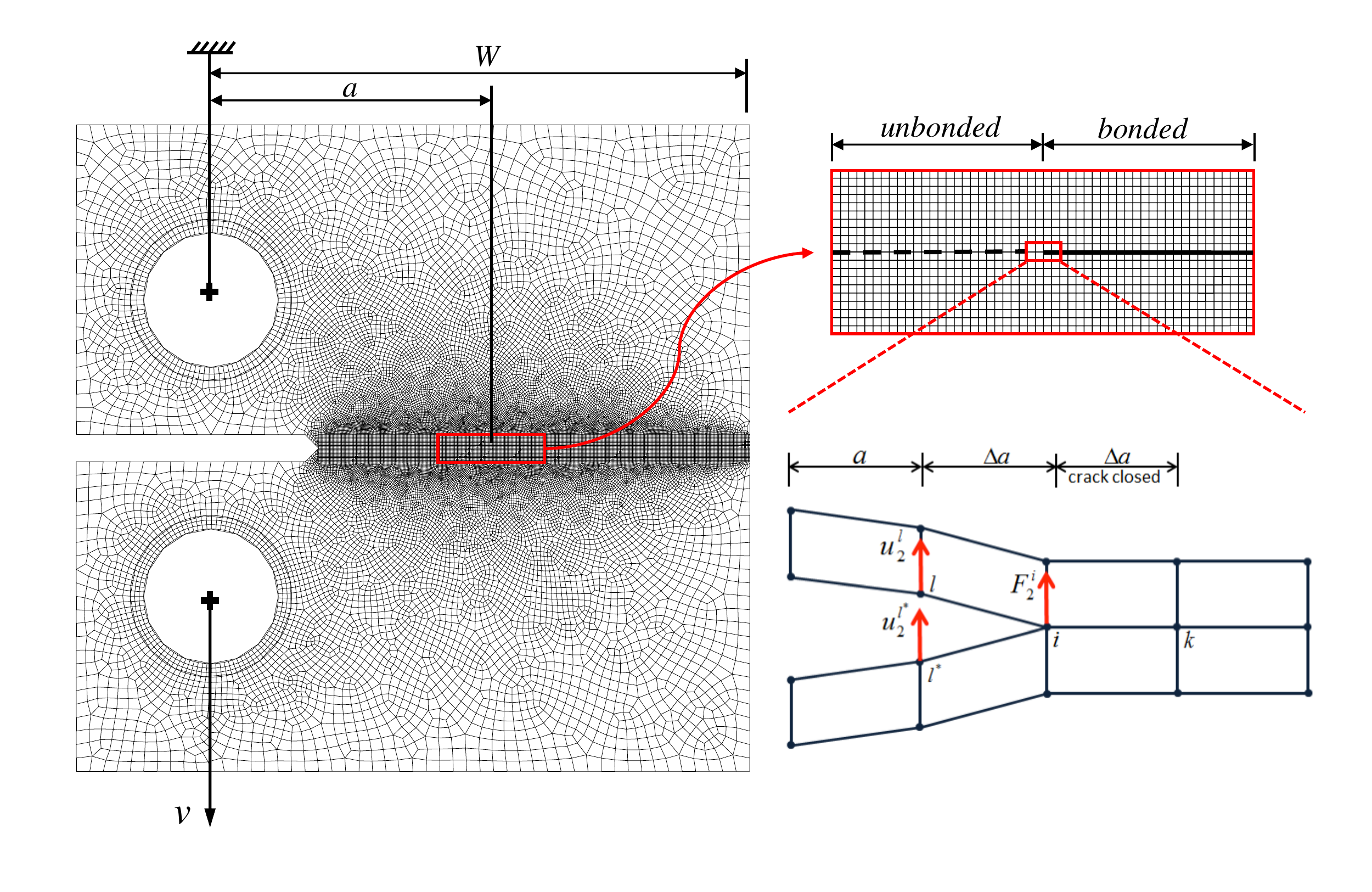}
\end{center}
\caption {Finite element boundary value problem, mesh geometry, and VCCT for four-noded elements.}
\vskip\baselineskip
\label{BVP-VCCT}
\end{figure}

In this study, a CT specimen is modeled in Abaqus finite element suite. The boundary value problem is shown in Fig.~\ref{BVP-VCCT}, where displacement of the top hole is fixed (but rotation is allowed) while a vertical displacement is applied to the bottom hole to mimic the Mode-I experiment. Isothermal loading path is adopted at $T=80^\circ$C, a temperature above $M_s$ and below $A_s$. Due to the chosen initial temperature, the entire material only undergoes stress-induced forward phase transformation and no reverse transformation takes place at any material point. Consequently, the material is initially in austenitic state and upon isothermal loading will undergo forward transformation (Fig.~\ref{PATH}). A pre-crack represented by ``unbonded" nodes (Fig.~\ref{BVP-VCCT}) is modeled that simulates the fatigue-induced pre-crack in a typical CT specimen. The ratio of the total crack size, \emph{a} (notch-plus-pre-crack), to specimen width, \emph{W}, is set to $a/W=0.5$ and crack propagates along a predefined path. A finite element mesh of four-node, isoparametric, quadrilateral elements is used which is highly refined along the crack line to accurately capture the near-tip fields (Fig.~\ref{BVP-VCCT}). Crack tip energy release rate ($G_I$) is used as the driving force for crack growth and virtual crack closure technique (VCCT) is implemented to calculate $G_I$. For the two-dimensional four-node elements placed in the crack front, the energy release rate is computed as
\begin{equation}\label{eq:G}
G_I=-\frac{1}{2 \Delta a}F_2^i(u_2^l-u_2^{l*}),
\end{equation}
where $F_2^i$ represent the nodal force at the crack tip and perpendicular to the crack plane, and $u_2^l$ and $u_2^{l*}$ indicate the opening displacement of the upper and lower crack surfaces, respectively (Fig.~\ref{BVP-VCCT}). Following assumptions are made: (i) a crack extension of $\Delta a$ does not significantly affect the state of the crack tip, (ii) the energy released during a crack extension of $\Delta a$ is identical to the energy required to close the crack by $\Delta a$. Crack growth occurs when $G_I$ attains the critical value, $G_{Ic}$. This critical value, \emph{i.e.} fracture toughness, is measured for a near-equiatomic NiTi at $T=80^\circ$C as $J_{Ic}=137~KJ/m^2$~\cite{haghgouyan2018experimental, haghgouyan2019fracture}.

\section*{RESULTS AND DISCUSSION}
\label{sec:results}

\begin{figure}[t]
\begin{center}
\includegraphics[width=1\columnwidth]{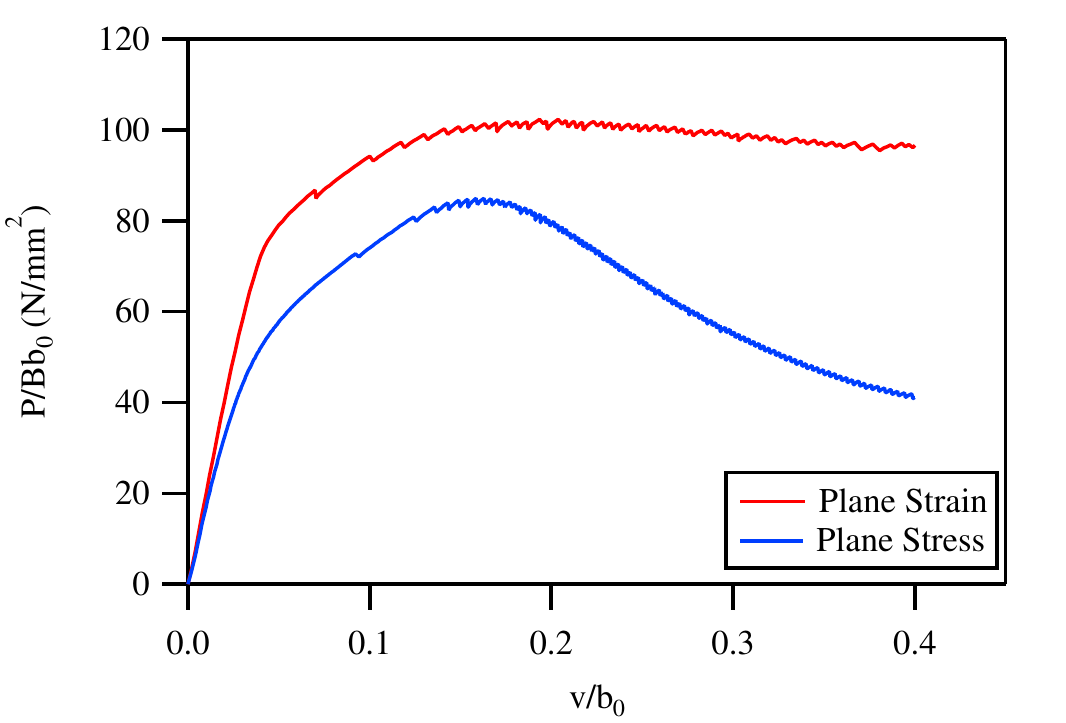}
\end{center}
\caption {Load per unit area of unbroken ligament vs. normalized displacement curves for isothermal loading of the CT SMA model under plane strain and plane stress conditions.}
\vskip\baselineskip
\label{pv}
\end{figure}

Numerically obtained load per unit area of unbroken ligament vs. normalized displacement curves under plane strain and plane stress isothermal tensile loading of the SMA CT specimen are shown in Fig.~\ref{pv}. Load is calculated from the reaction force at the top pin hole while displacement is obtained from the bottom pin hole, where the boundary condition is imposed. The load-displacement curves in plane strain and plain stress are characterized by a linear response in the early stages of loading (corresponding to linear elastic response of pure austenite), followed by a deviation from linearity (corresponding to forward phase transformation from austenite to martensite starting at the crack tip) and a maximum tensile load, followed by gradual load drop. While the general trends in the two curves are alike, there are some key differences. After a linear elastic response in the beginning, the curve corresponding to plane stress condition deviates from linearity earlier compared to plane strain. The load drop after is also much more pronounced under plane stress. These observations can be explained based on the stress state that exists in the material in these two cases: under plane strain, an out-of-plane stress component exists in the specimen and thus the equivalent stress at the crack-tip is lower compared to that under plane stress.

\begin{figure}[t]
\begin{center}
\includegraphics[width=0.9\columnwidth]{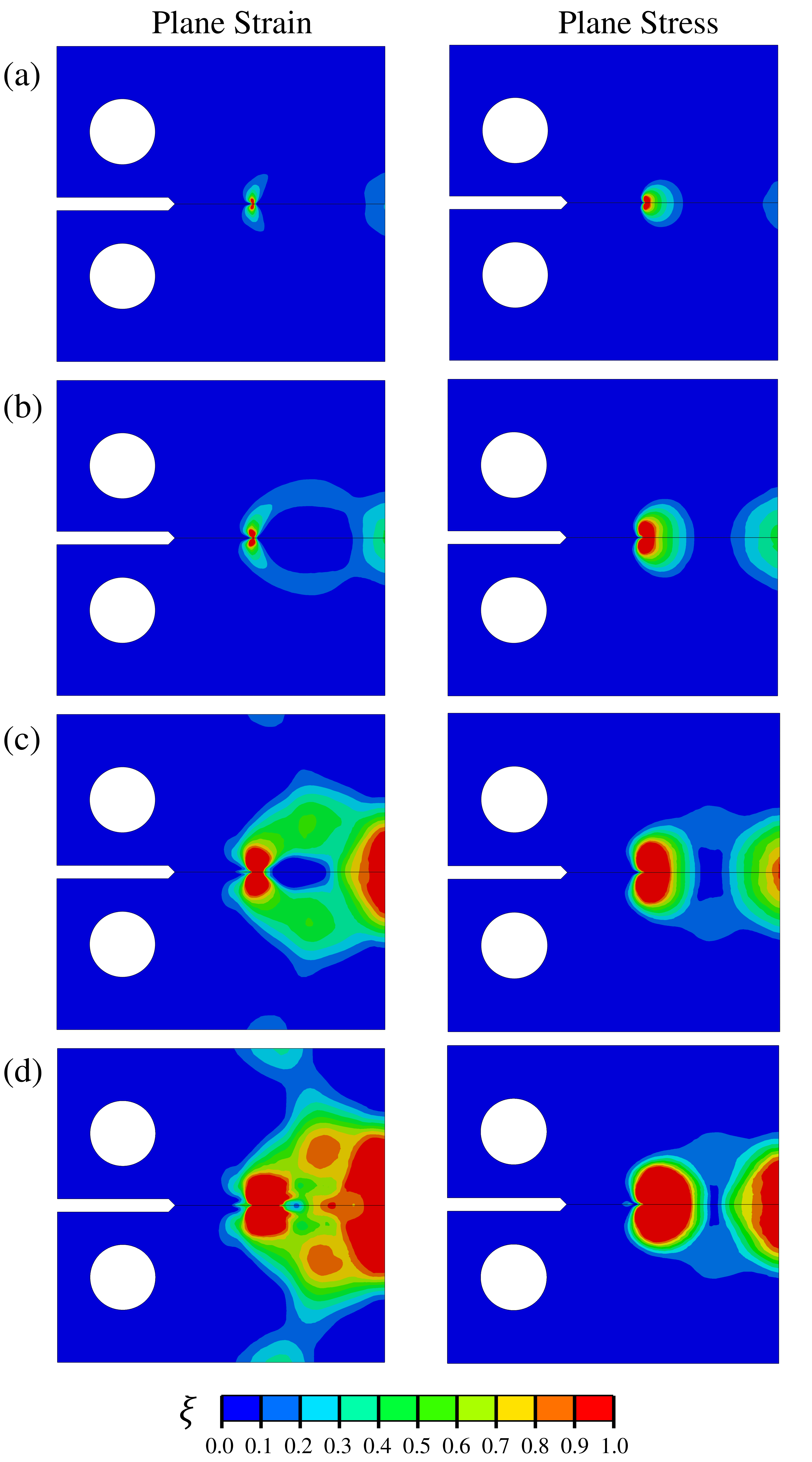}
\end{center}
\caption {Evolution of martensite volume fraction, $\xi$, for plane strain (left) and plane stress (right) analyses: (a) at the early stages of loading corresponding to a vertical displacement of $v=0.5~mm$; (b) at the onset of crack extension; (c) at the maximum attained load; and (d) after $2~mm$ crack extension.}
\vskip\baselineskip
\label{mvf}
\end{figure}

The difference between load-displacement response of plane strain and plane stress cases can be further explained by examining the stress-induced martensitic transformation zone. Upon mode-I loading, the stress at the crack tip increases and stress-induced transformation zone appears when it reaches the critical value for transformation. To that end, the evolution of martensite volume fraction, $\xi$, is plotted at the following points: (i) at the early stage of loading corresponding to a vertical displacement of $v=0.5~mm$, (ii) at the onset of crack extension, (iii) at the maximum attained load, and (iv) after $2~mm$ crack extension. This is presented in Fig.~\ref{mvf} (a),(b),(c), and (d), respectively, for plane strain (left) and plane stress (right) analysis. Martensite volume fraction contours are mapped onto the reference configuration of the material specimen for ease of comparison. Comparing the plane strain and plane stress results, in Fig.~\ref{mvf}(a), the region of fully transformed material is relatively larger and transformation occurs primarily in front of the crack tip in plane stress, as apposed to that in plane strain, where the transformation zone extends along two lobes at an angle to the crack line on either sides. Since the out of plane stress component is zero in plane stress, the maximum shear stress occurs in planes making $45^{\circ}$ with the plane of the specimen. On the other hand, the maximum shear stress in plane strain occurs in planes normal to the plane of specimen making $45^{\circ}$ with the crack plane. Consequently, the stress required to initiate transformation at a material point close to the crack-tip is higher in plane strain, giving rise to a smaller transformation zone than that in plane stress~\cite{gdoutos2006fracture}. At the onset of crack extension, Fig.~\ref{mvf}(b), the transformation zone expands in both cases, while in plane strain it starts to interact with the transforming region under compression at the opposite end of the specimen. When the energy release rate at the crack tip reaches the critical value, the crack extends into a region of fully transformed martensite. At maximum load (Fig.~\ref{mvf}(c)), interaction between the near-tip transformation and the transformed region at the end of the specimen is also visible in plane stress. In the plane strain maximum load case, the transformation zone is extended well beyond the near-tip region. After substantial crack propagation (Fig.~\ref{mvf}(d)), bulk of the material in front of the crack tip has transformed in plane strain and transformation zone extends to the specimen boundaries; whereas under plane stress, the near-tip and compressive transformation region has grown in size.

\begin{figure}[t]
\begin{center}
\includegraphics[width=1\columnwidth]{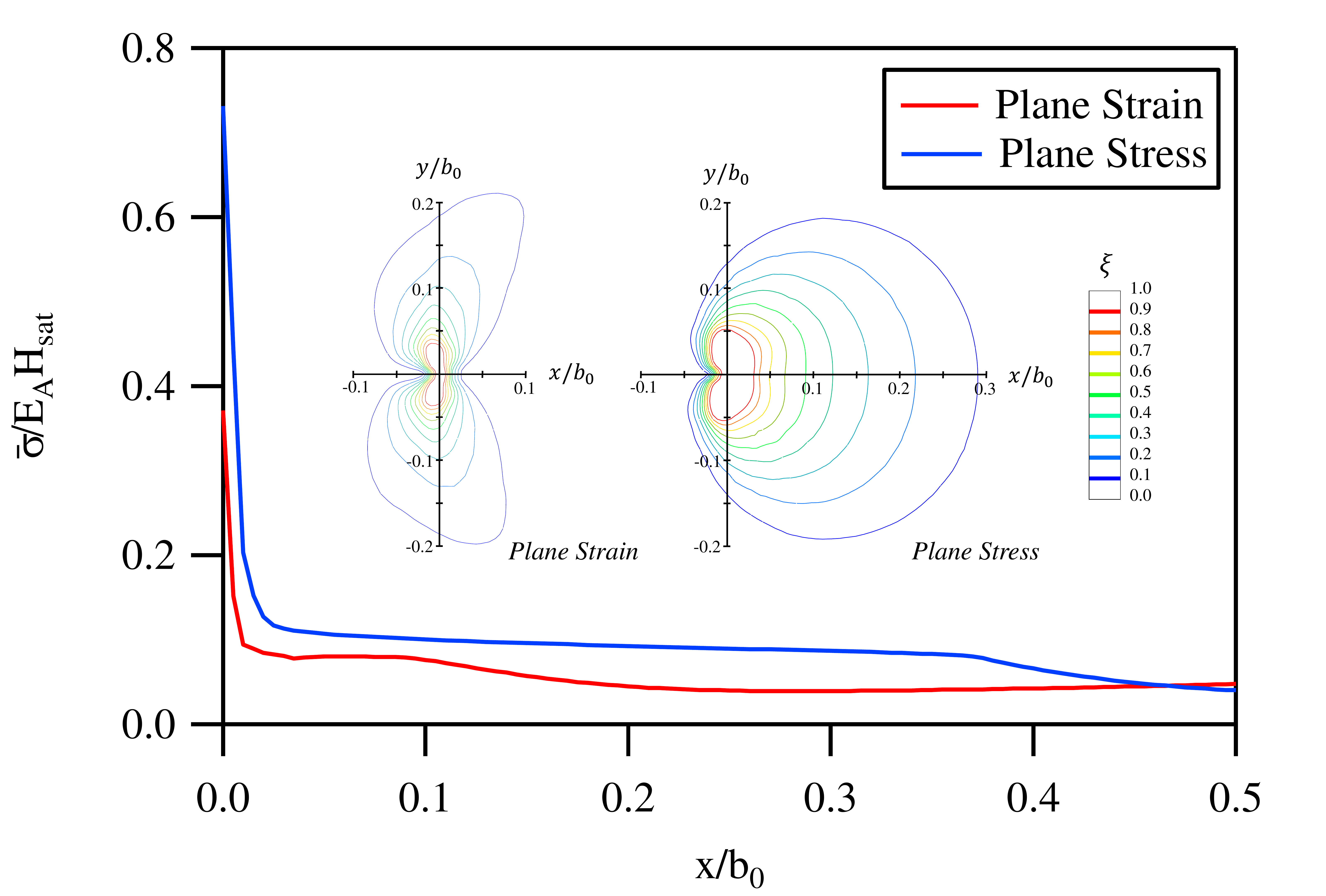}
\end{center}
\caption {Stress distribution near the crack tip of CT SMA model under plane strain and plane stress conditions at $v=0.5~mm$ along with the corresponding transformation zone boundaries.}
\vskip\baselineskip
\label{stress}
\end{figure}

Normalized von Mises equivalent stress, $\bar{\sigma}$, ahead of the crack-tip for plane strain and plane stress conditions prior to crack extension ($\emph{i.e.}$ when $v=0.5~mm$) is shown in Fig.~\ref{stress}. The stress is normalized by $E_A H_{sat}$ and the length parameter is normalized by initial length of the unbroken ligament, $b_0$. The stress distribution can be explained by considering three regions at the vicinity of crack-tip: a fully transformed region close to the tip, an untransformed region further away from it, and a partially transformed region in between. Inside the untransformed austenite, stress increases as one approaches the crack tip. A plateau is observed in the partially transformed region which contains a mixture of the austenite and martensite phases. Stress increases abruptly within the fully transformed martensite in the region very close to the crack tip. The corresponding transformation zone boundaries are also included in the figure to better understand the role of stress-induced transformation. Although the overall variation of stress is similar, magnitude of the von Mises stress is higher under plane stress than that under plane strain.

\begin{figure}[t]
\begin{center}
\includegraphics[width=1\columnwidth]{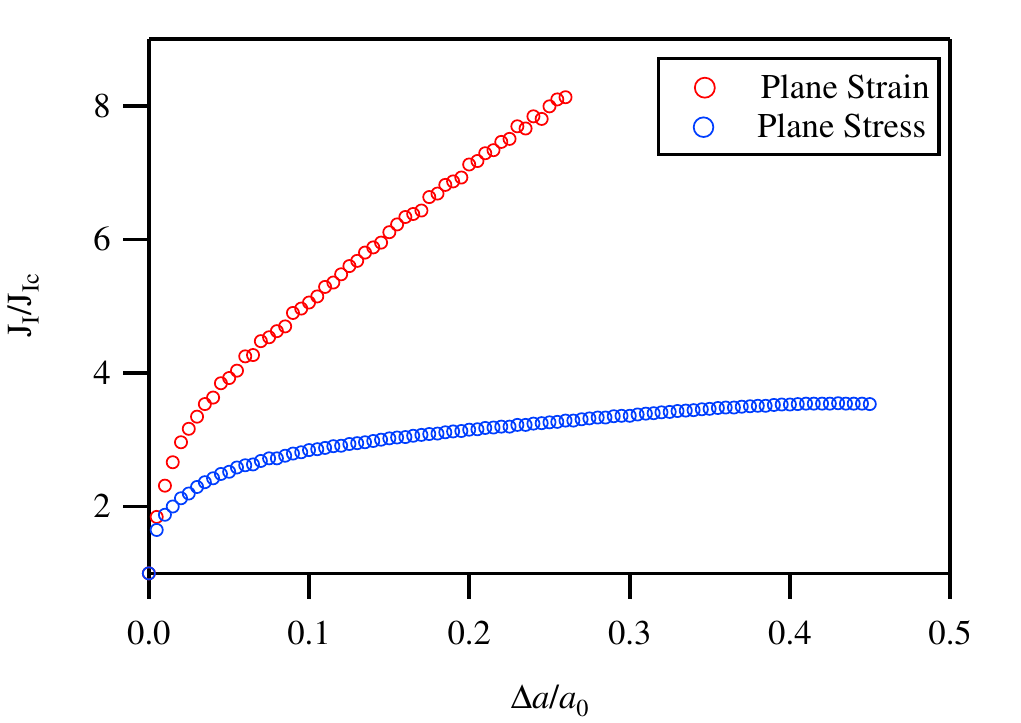}
\end{center}
\caption {Resistance curves for isothermal loading of the CT SMA model under plane strain and plane stress conditions.}
\vskip\baselineskip
\label{r-curve}
\end{figure}

Crack growth resistance curve (R-curve) under plane strain and plane stress are plotted in Fig.~\ref{r-curve}. At each point corresponding to a node release in the finite element mesh, $G_I$ is calculated from the load-displacement curve and normalized by $G_{Ic}$-value corresponding to the onset of crack extension. Crack extension, $\Delta a$, is obtained by post processing the status of nodes along the predefined crack path, and is normalized by the the initial crack size, $a_0$. A rising R-curve is observed for both plane stress and plane strain conditions. The rising behavior indicates fracture toughness enhancement and is associated with closure stresses acting on the crack tip as a result of the transformed material left behind in the wake of the advancing crack tip~\cite{baxevanis2013fracture}. In other words, as the crack extends into the fully transformed material, the martensitic material left behind the crack and provides shielding to the crack that leads to an apparent increase in the fracture toughness, requiring further loading to sustain the stable crack extension. Out of the total energy supplied to the specimen from applied external loading, a part is utilized in transformation dissipation while the rest is separation work, \emph{i.e.} work required to create new crack surfaces. Due to these competing phenomena, as long as the energy dissipated due to phase transformation is relatively small compared to separation work, crack propagation will dominate as a mechanism for energy dissipation and crack will grow in the fully transformed martensitic region in a quasi-static stable manner. Comparing the plane strain and plane stress results, it is obvious that slope of the R-curve is more gradual for the latter analysis, and is apparent in the pronounced softening during load drop in the associated load-displacement behavior (Fig.~\ref{pv}). This softening points to a preference to crack propagation as opposed to phase transformation as a mechanism for dissipation under plane stress. The contribution of transformation strain work is small compared to the separation work in plane stress condition, since a large region close to the crack tip is already fully transformed.  Plane strain crack resistance curve, on the other hand, shows a steep increase in the crack tip energy release rate during crack propagation. This indicates that increasing amount of energy supplied to the material in plane strain is utilized in transformation dissipation rather than formation of new crack surfaces, and is apparent in the generation of relatively large scale transformation regions in plane strain (Fig.~\ref{mvf}(d)).

\section*{CONCLUSIONS}
\label{sec:conclusion}
In this work, crack growth problem in an SMA CT specimen subjected to mode-I, isothermal loading is studied in Abaqus finite element suite. VCCT is used to model crack propagation and crack growth is investigated under plane strain and plane stress conditions. The load-displacement response is obtained for both cases and the differences in characteristics of the curves are explained. Transformation zone near the crack-tip is also studied by examining the evolution of martensite volume fraction during different stages of loading and resultant crack growth, and it is found that the transformation zone is larger and more pronounced at the crack-tip region under plane stress. The effect of transformation zone on stress distribution ahead of the crack-tip is discussed. Moreover, the crack resistance R-curve is obtained under plane strain and plane stress conditions and a rising R-curve is observed in both cases indicating fracture toughness enhancement. It is found that slope of the R-curve is more gradual for the plane stress analysis due to softening in the load-displacement curve whereas a much steeper response is observed under plane strain, and the different behavior is explained in terms of the dominant mechanism for energy dissipation in each case: phase transformation in plane strain vs. quasi-static stable crack propagation in plane stress.
\bibliographystyle{asmems4}

\begin{acknowledgment}
This material is based upon work supported by AFOSR under Grant No.~FA9550-15-1-0287.
\end{acknowledgment}

\bibliography{smasis}

\end{document}